\documentclass[journal]{IEEEtran}
\usepackage{graphicx}
\usepackage{subcaption}

\usepackage{amsmath}
\usepackage{soul,color}
\usepackage{float}
\usepackage{dirtytalk}
\usepackage[nolist,nohyperlinks]{acronym}
\usepackage{algorithm}
\usepackage[noend]{algpseudocode}
\algblockdefx{MRepeat}{EndRepeat}{\textbf{repeat}}{}
\algnotext{EndRepeat}
\usepackage{amsthm}
\usepackage{epstopdf}
\usepackage{tikz}

\newtheorem{definition}{Definition}

\usepackage{cite}

\ifCLASSINFOpdf
\else
\fi
\begin{document}
%
\title{Agent-Based Modelling Approach for Distributed Decision Support in an IoT Network}
%
%
%
\author{Merim~Dzaferagic,
  M. Majid Butt,~\IEEEmembership{Senior~Member,~IEEE,~}Maria Murphy,
         Nicholas~Kaminski,
        and~Nicola~Marchetti,~\IEEEmembership{Senior~Member,~IEEE}

        \thanks{Merim~Dzaferagic Maria Murphy, Nicholas~Kaminski and Nicola~Marchetti are with CONNECT center for future networks, Trinity College Dublin, Ireland. Email:\{dzaferam, nicola.marchetti\}@tcd.ie.}
\thanks{M. Majid Butt is with School of Engineering, University of Glasgow, UK. Email: majid.butt@glasgow.ac.uk.}
\thanks{
This publication has emanated from research conducted with the financial
support of Science Foundation Ireland (SFI) and is co-funded under the European
Regional Development Fund under Grant Number 13/RC/2077.}
}

\maketitle
\begin{abstract}
An increasing number of emerging applications, e.g., internet of things, vehicular communications, augmented reality, and the growing complexity due to the interoperability requirements of these systems, lead to the need to change the tools used for the modeling and analysis of those networks. Agent-Based Modeling (ABM) as a bottom-up modeling approach considers a network of autonomous agents interacting with each other, and therefore represents an ideal framework to comprehend the interactions of heterogeneous nodes in a complex environment. Here, we investigate the suitability of ABM to model the communication aspects of a road traffic management system, as an example of an Internet of Things (IoT) network. We model, analyze and compare various Medium Access Control (MAC) layer protocols for two different scenarios, namely uncoordinated and coordinated. Besides, we model the scheduling mechanisms for the coordinated scenario as a high level MAC protocol by using three different approaches: Centralized Decision Maker, \textsc{Desync} and decentralized learning MAC (L-MAC). The results clearly show the importance of coordination between multiple decision makers in order to improve the accuracy of information and spectrum utilization of the system. 
\end{abstract}


\section{Introduction} \label{sec:introduction}
In the context of this study, a complex system is defined as \textit{any system
  featuring a large number of interacting components (agents, processes, etc.)
  whose aggregate activity is nonlinear (not derivable from the summations of
  the activity of individual components) and typically exhibits hierarchical
  self-organization under selective pressures} \cite{grady2010system}.
Considering this definition of complex systems, the next generation of communication
networks (e.g., \ac{IoT}, cellular networks, vehicular networks) can be regarded as a complex system due to growing number of
technologies and connected devices. Complexity in decision making (scheduling, routing) for
a large IoT system requires new modeling and decision making tools and methodologies, which motivates the study of complex systems science (CCS)
\cite{6883525,Macaluso2016}. The tools used to model and analyze these networks
must evolve in order to optimally utilize the available resources (e.g.,
spectrum, processing power) at affordable complexity.

\ac{ABM} is a bottom-up method of modeling that considers a network of
autonomous agents. Each agent has its own set of attributes and behaviors. These
behaviors describe how the agents interact with other agents and their
environment. If needed, the agents can exhibit learning capabilities that allow
them to adapt to changes in the system, altering the internal attributes and the
behaviors towards other agents. Therefore, \ac{ABM} is suitable to model complex
systems \cite{Macal, 4117593, wilensky2015introduction} that would require large
computational complexity to be modeled otherwise.

\ac{ABM} has previously been used to model a wide range of applications in
sectors such as ecology, biology, telecommunications and traffic management.
Some examples include: \cite{Pogson200637} where \ac{ABM} is used to model
intra-cellular chemical interactions, and \cite{Benenson2008} to analyze the
parking behaviors in a city.

Recently, ABM has been used to solve various complex problems in
telecommunication networks. The authors of \cite{laghari2016modeling} show that
a cognitive agent-based computing modeling approach, such as \ac{ABM}, is an
effective approach to model complex problems in the domain of \ac{IoT}. Our
work examines the use of \ac{ABM} to model an \ac{IoT} network that requires
distributed decisions. The \ac{IoT} network in question is a road traffic
management system that adjusts the timing of traffic lights based on the amount
of vehicles waiting at an intersection. Our focus is on the modeling of the
communication aspects of the system and in particular we want to analyze the
impact of the \ac{MAC} protocol on the application itself, i.e., the timing of
traffic lights.

Many applications of \ac{ABM} in the telecommunications industry have focused on
economic and social aspects, such as consumer behavior. In
\cite{twomey2002agent}, the authors model the customer behavior in a
telecommunication network. Also, in \cite{12029839}, \ac{ABM} is used to analyze
the wireless cellular services market. There have been quite a few applications of
\ac{ABM} that model the network itself. In \cite{pittir9587}, the authors
describe how \ac{ABM} can be applied to model spectrum sharing techniques in
future 5G networks. The authors model a system that considers economical,
technical and regulatory considerations when leasing spectrum. The agents are
able to remember what spectrum sharing conditions were beneficial for them
previously and learn/adapt based on their previous choices. The authors of
\cite{horvath2013agent} analyze the spectrum frequency trading mechanism by
modeling the heterogeneous nodes as agents in an \ac{ABM} framework. The
motivation for \ac{ABM} came from the emergence of structures, patterns and
unexpected properties. \ac{ABM} allowed them to model and understand market
models with dynamics that are beyond the scope of familiar analytical
formulations, such as differential equations. 

Other applications of \ac{ABM} to communication networks are presented in
\cite{niazi2009agent, liu2016decentralized, niazi2011novel}. The authors of
\cite{niazi2009agent} analyze the effectiveness of \ac{ABM} to model
self-organization in peer-to-peer and ad-hoc networks. They also outline the
limitations of using current modeling and simulation software. Their work shows
that tools such as OMNET++, Opnet and specialized tools such as the Tiny OS
Simulator are limited as they tend to focus solely on computer networks.
Interactions with humans and mobility cannot be modeled with enough flexibility.
Network parameters can be easily modified but other conditions are difficult to
be considered. The authors highlight the flexibility of an \ac{ABM} approach,
showing how easily the system can be updated and allow for powerful result
abstraction. In \cite{liu2016decentralized}, the authors analyze a decentralized
spectrum resource access model as a complex system, modeling the decentralized
decision making and cooperation of distributed agents in a way that allows them
to partially observe the state of the system, meaning that each agent has only
the information about its own local environment. The authors of
\cite{niazi2011novel} introduce an \ac{ABM} framework to formally define all
necessary elements to model and simulate a \ac{WSN}. As a proof of concept, they
demonstrate the application of the framework to a model of self-organized
flocking of animals monitored by a random deployment of proximity sensors. By
proving the applicability of \ac{ABM} to \ac{WSN}s, this paper provides further
motivation to the current study.

Many studies have been carried out to try to optimize traffic flow (specifically
in urban areas) using \ac{WSN}. For example, in \cite{coleri2004sensor},
\cite{tubaishat2007adaptive} and \cite{cheung2005traffic}, sensor networks for
monitoring traffic are proposed. The authors of \cite{Ma2016164, Hager2015306,
  lansdowne2006traffic} use \ac{ABM} for traffic optimization and simulation.
Reference \cite{Ma2016164} describes an \ac{ABM} solution to generate personalized
real-time data to present route information to travelers. The authors in \cite{Hager2015306}
model the effect of an increasing population on traffic congestion. In
\cite{lansdowne2006traffic} a detailed traffic simulator using NetLogo was
designed. It analyzes the effect on traffic congestion when various different
lanes of traffic are introduced. Though, the
work presented in those papers provide motivation to our work, their focus lies
in the functionality and optimization of the traffic light systems and not on
the modeling and analysis of the communication aspects of the related sensor
network.

Due to the suitability of the \ac{ABM} approach to model large systems composed
of autonomous decision making entities, we believe that \ac{ABM} is a perfect
match to model and analyze the problem addressed by this paper, i.e., road
traffic management system. \ac{ABM} allows us to model individual agents and the
effect that those agents have on their local environment, and as a result we can
observe the cumulative/system level behavior that results from the agents
interacting with each other and with the environment. The beauty of this
approach is that by modeling the interactions and their effects locally, we
actually model a complex decision making system which is decentralized in
nature. It may not be optimal as compared to centralized decision making
entities, but it provides a low complexity decision making framework.

Building on the preliminary work in \cite{kaminski2016_maria}, we have proposed
a comprehensive analysis of an ABM approach to model a traffic intersection
system. The main contributions of this article are summarized as:
\begin{itemize}
\item We demonstrate the use of \ac{ABM} to model the
  communication aspects of \ac{IoT} networks;
\item We investigate the impact of different \ac{MAC} protocols on the accuracy
  of information gathered by the sensor network;
\item Using ABM, we evaluate the spectrum utilization of different MAC protocols (i.e.,
  TDMA, slotted Aloha and CSMA/CA) in a multi-layered network configuration;
\end{itemize}

The rest of the paper is organized as follows. Section
\ref{sec:model_description} presents the description of the model, and outlines
the algorithms and \ac{MAC} protocols that are implemented in the \textit{Mesa}
framework. In Section \ref{sec:analysis}, we present the methodology used for
the analysis and discuss the results gathered from the simulations. In Section
\ref{sec:conclusion}, we elaborate on the main findings, and draw the overall
conclusions on the work.

\section{Description of Model} \label{sec:model_description}
\subsection{Agent Based Modeling Framework}
We first define the fundamental terms used in ABM.
\begin{definition}[Agent]
An autonomous computational object with particular properties and capable of
particular actions is called an agent.
\end{definition}
\begin{figure}[t]
  \centering
  \centerline{\includegraphics[scale=0.3]{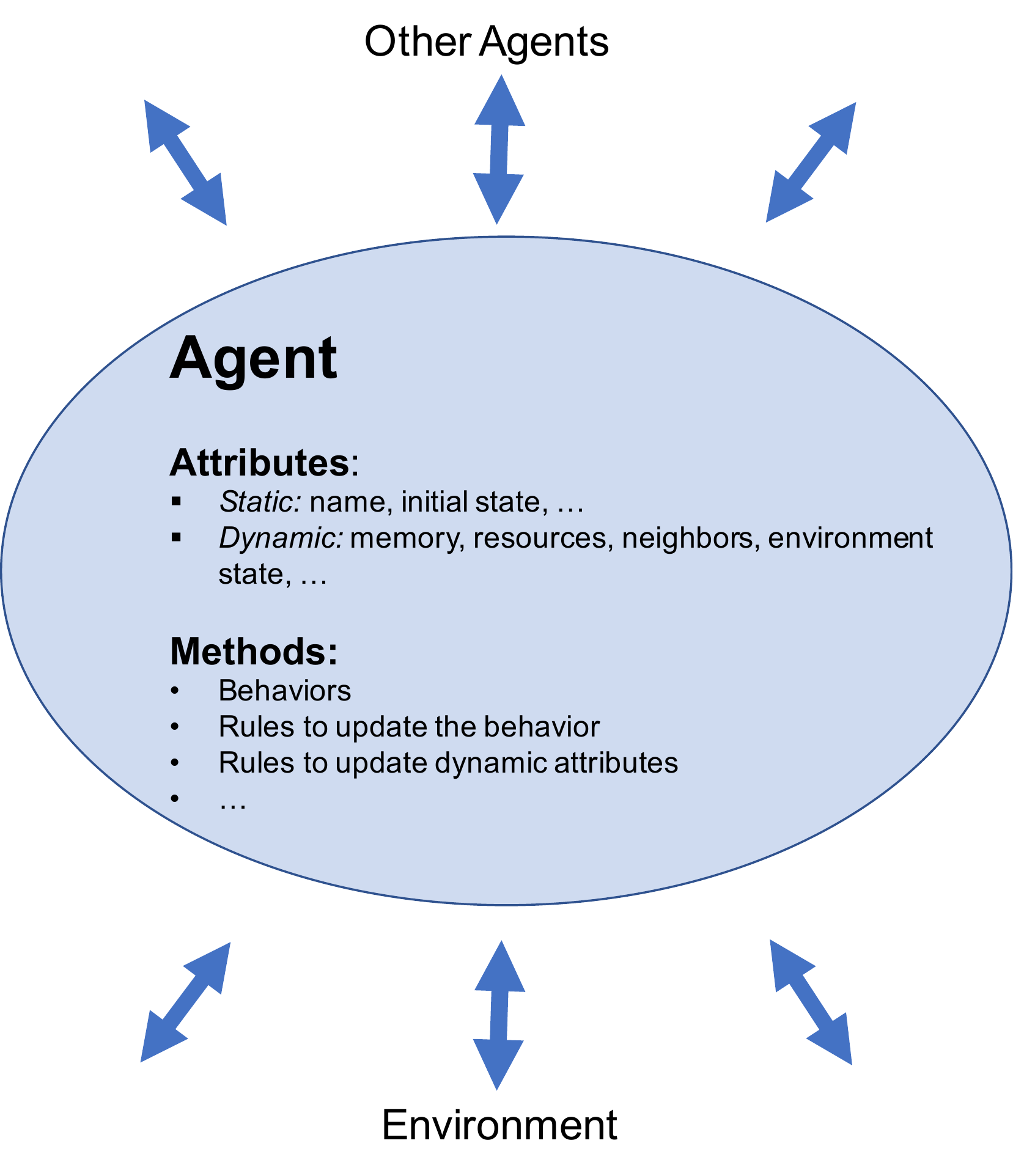}}
  \caption{Each agent can interact with the environment or other neighbors in
    its neighborhood. Agents also have a set of static and dynamic attributes
    storing the properties of the agent and its knowledge about the surrounding
    agents and the environment. The simple rules that an agent is following are
    encoded in the methods.}
  \label{fig:agent_description}
  \vspace{-1.5em}
\end{figure}
Agents are completely autonomous entities in their decision making. As shown in
Figure \ref{fig:agent_description}, every agent has a set of attributes and
methods that define how and with whom it can interact. This defines the
topology of an ABM system. Not all the agents are connected with each other;
instead an agent is connected with a particular set of agents, called
\textit{neighbors}, who influence its localized decision making.
\begin{definition}[Environment]
A set of entities that influence the behavior of an agent constitute the
environment for an agent.
\end{definition}
Figure \ref{fig:agent_description} shows that agents interact not only with
other agents but also with the environment.

To model a problem using \ac{ABM}, we have to define the agents, the environment
as well as associated methods and interactions in a way that reflects the
original problem. It should be noted that the goal of our \ac{ABM} approach is
not to optimize the system, but to model the system as accurately as possible.
The goal is to model the distributed optimization mechanisms that would converge reasonably to optimized solution by modeling the interactions of a
decentralized system such that the complexity remains manageable.

\subsection{Road Traffic ABM Model}
\begin{figure}[t]
  \centering
  \centerline{\includegraphics[scale=0.35]{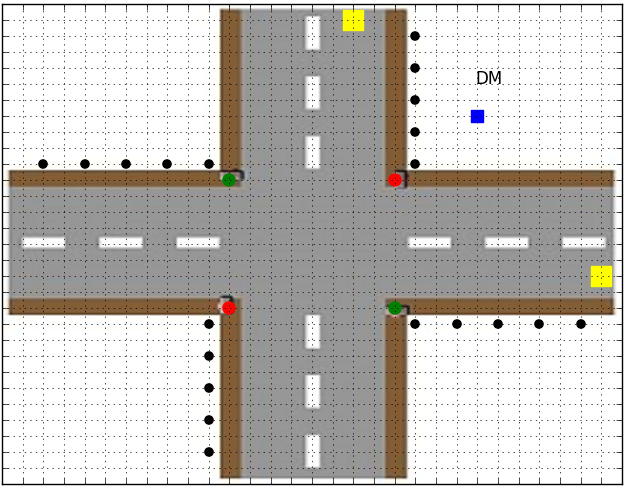}}
  \caption{The sensors are represented by the black dots surrounding the
    perimeter of the road. The green and red dots are traffic lights. The
    vehicles are represented by yellow squares, and the \ac{DM} is represented
    by a blue square in the upper right section.}
  \label{fig:oneIntersection}
  \vspace{-1.5em}
\end{figure}

In order to explain the model (i.e., the agents and the environment
of the \ac{ABM} model), we start with the single intersection of roads. The
visualization of this model is shown in Figure \ref{fig:oneIntersection}. The
environment of our model is represented by the road. The agents in our model
are:
\begin{itemize}
\item sensors
\item traffic lights
\item vehicles
\item \ac{DM}
\end{itemize}
This single intersection model contains 20 sensors. The sensors are represented
by the black dots surrounding the perimeter of the roads. There are four traffic
lights. These can be seen as the red/green dots near the center of the image.
The yellow squares in Figure \ref{fig:oneIntersection} symbolize vehicles
traveling on the road.\footnote{It should be noted that the vehicles follow UK and
Ireland driving conventions and travel on the left-hand side of the road.} The
blue square in the upper right section of the image represents a central
\ac{DM} that will be responsible for managing the timing of the traffic lights
in the model.

The interaction between the sensor agents and the environment (the road) is
modeled by the collection of traffic measurements. Those measurements represent
the number of vehicles approaching the intersection. Once a sensor observes a
vehicle approaching the intersection it tries to transmit this information to
the \ac{DM}.

The \ac{DM} controls the timing of the four traffic lights with the aim of
optimizing the waiting time of the vehicles traveling through the intersection.
It is important to notice that we do not focus on the optimization algorithms
related to the vehicle traveling time, we rather focus on the analysis of the
communication aspects of the sensor network that collects the information about
the traffic. In Figure \ref{fig:oneIntersection}, the traffic lights appear red
and green in colour. The red colour symbolises that traffic should stop when it
reaches the traffic light. The green colour represents that the vehicles are
free to move past the traffic light.

\begin{algorithm}[t]
\caption{Model}\label{alg:model_flow}
\begin{algorithmic}[0] 
  \State $T_{c} \gets \textit{Current Tick number}$
  \State $T_{max} \gets \textit{Tick number limit}$
  \MRepeat
  \State $\textit{Set tick\_number to 0} $
  \For {$\textit{each existing vehicle}$}
  \If {$\textit{no car in front and green light}$}
  \State $\textit{Move forward}$
  \EndIf
  \EndFor
  \If {$\textit{random\_number $<$ probability of new vehicle}$}
  \State $\textit{Create new vehicle}$
  \EndIf
  \While {$\textit{$T_c < T_{max}$ \textbf{or} Transmission not succ.}$}
  \State $\textit{Sensor attempt to transmit}$
  \State $\textit{Increment $T_c$}$
  \EndWhile
  \If {$\textit{$T_c == T_{max}$}$}
  \State $\textit{DM make decision}$
  \EndIf
  \EndRepeat
\end{algorithmic}
\end{algorithm}

The model description is outlined with the Algorithm \ref{alg:model_flow}. The
tick number limit ($T_{max}$) is a user defined parameter, that allows us to
define the upper time limit for the sensor transmission attempts and at the same
time the decision making interval of the \ac{DM} agent. One tick is equivalent
to a transmission time slot on the \ac{MAC} layer. Since we dedicate one tick to
the movement and generation of the vehicles and one additional tick for the
\ac{DM} decision making function, the number of ticks that is dedicated for the
transmission of the measurements ($T_{trans}$) is calculated as:

\begin{equation}
  \label{eq:transmission_interval}
  T_{trans} = T_{max} - 2
\end{equation}

As shown in Algorithm \ref{alg:model_flow}, each car that is currently on the
grid will attempt to move one space forward in each simulation iteration (one
iteration takes $T_{max}$ ticks). For simplicity sake, the vehicles will always
travel in a straight line. If there is already another vehicle in the space a
certain car intends to move to or the space in front of that, it will not be
allowed to move forward. This prevents vehicles from colliding with each other
or travelling too close to each other. Vehicles are also prohibited from moving
if they are close to a traffic light in their trajectory and the traffic light
is red.

Each simulation iteration involves the creation of new vehicles on the
grid. The number of vehicles added to the grid depends on the user defined
parameter ($\textit{prob. of new vehicle}$). Vehicles will only be initialized
on the edges of the grid either traveling north, south, east or west. The
initial position of the newly generated vehicles is chosen randomly (i.e.,
uniformly sampled from a list of all available positions). If there is already a
vehicle currently blocking the placement of the newly generated vehicle, the
newly generated vehicle will be discarded.
Once the vehicles are placed on the grid and the existing vehicles move
according to the above mentioned rules, the sensors collect the vehicle
position information and attempt to convey those measurements to the
\ac{DM}. The sensors detect stopped vehicles by remembering the grid space where
vehicles were detected in the previous cycle. If this grid space is still
occupied by the same vehicle in the current cycle, that means that the vehicle
has stopped moving. On the other hand, if the grid space is no longer occupied,
the vehicle has moved on. We model the \ac{MAC} layer protocols (i.e., TDMA,
slotted Aloha and CSMA/CA) for the communication between the sensors and the
\ac{DM}s, and in the case of multiple intersections we also model the
communication between the \ac{DM}s (TDMA like scheduling). It is important to
keep in mind that the implemented model is discrete (the time is divided into
slots of equal duration). Each sensor can only transmit at the beginning of a
slot. The chosen \ac{MAC} protocol defines how to deal with potential
collisions. A collision happens in case a sensor attempts to transmit in the
same slot as one of its neighbors. A neighboring sensor is the one that is in the
selected sensor's Moore neighborhood. The Moore neighborhood represents the 8
grid spaces surrounding the selected sensor's grid space. Hence, a sensor
transmission is affected by the transmission of sensors in all directions,
including diagonals.

\begin{figure}[t]
  \centering
  \centerline{\includegraphics[scale=0.35]{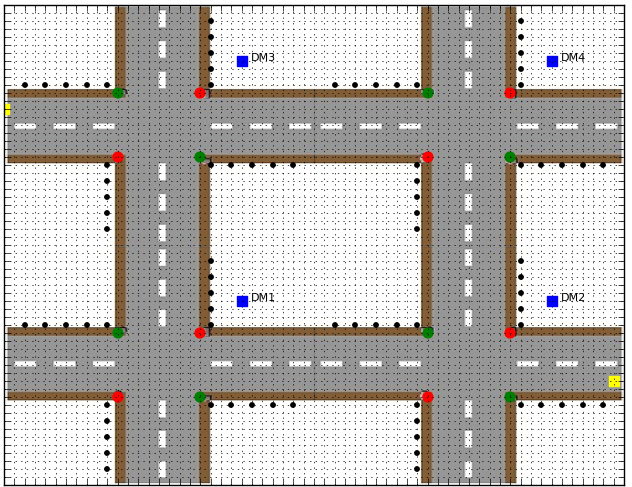}}
\caption{A four intersections model, showing all the sensors (black), traffic
  lights (red and green), vehicles (yellow) and decision makers (blue) that are
  part of our model.}
\label{fig:four_intersection_model}
\vspace{-1.5em}
\end{figure}

In our analysis we consider three \ac{MAC} protocols for the communication
between the sensor nodes and the \ac{DM}s, i.e., slotted Aloha, TDMA and
CSMA/CA. Slotted Aloha deals with collisions by introducing back-off time,
meaning that in case two neighboring sensors try to transmit at the same time,
both transmissions will be unsuccessful and the sensors will choose a random
back-off time to retry the transmission. The Aloha protocol also introduces a
timeout time, which in case it is reached without a successful transmission
implies that the packet should be discarded. As opposed to the Aloha protocol
which does not involve any type of synchronization between the nodes and
therefore, potentially leads to packet collisions, the TDMA protocol is
implemented by allowing the \ac{DM} to assign each sensor a specific time slot
for packet transmissions. The centralized coordination, results in a collision
free environment, if only one \ac{DM} exists (i.e., the single intersection of
roads scenario). In case multiple \ac{DM}s are managing the communication of
their sensors (i.e., Figure \ref{fig:four_intersection_model}), we have to
introduce some type of coordination between the \ac{DM}s to avoid potential
packet collisions. The CSMA/CA protocol, like Aloha, is an opportunistic
approach. If a sensor has information to send, it first checks if any of its
neighbours is currently using the spectrum resource. If the resource is
currently being used, a back-off time will be computed. If the resource is not
being used, the packet will be transmitted collision free and the \ac{DM} will
send an acknowledgment packet back to the sensor node to confirm the reception
of the packet. It should be noted that the model assumes that there are no
hidden nodes. Therefore, all potential collisions will be successfully sensed
before transmission.

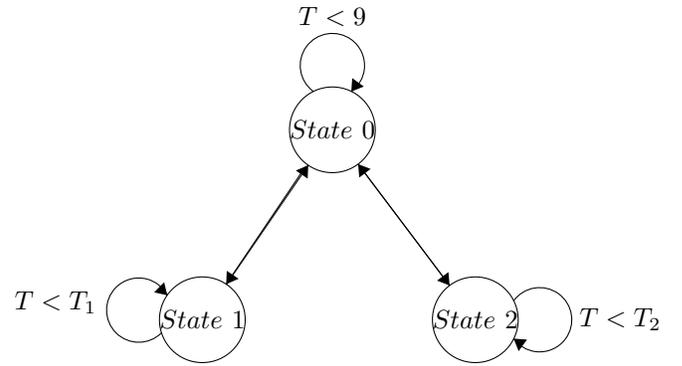
\begin{figure}[t]
  \begin{center}
    \begin{tikzpicture}[scale=0.19]
      \tikzstyle{every node}+=[inner sep=0pt]
      \draw [black] (36.5,-12.2) circle (3);
      \draw (36.5,-12.2) node {$State\mbox{ }0$};
      \draw [black] (27.4,-25.5) circle (3);
      \draw (27.4,-25.5) node {$State\mbox{ }1$};
      \draw [black] (46.5,-25.5) circle (3);
      \draw (46.5,-25.5) node {$State\mbox{ }2$};
      \draw [black] (38.3,-14.6) -- (44.7,-23.1);
      \fill [black] (44.7,-23.1) -- (44.62,-22.16) -- (43.82,-22.76);
      \draw [black] (49.18,-24.177) arc (144:-144:2.25);
      \draw (53.75,-25.5) node [right] {$T<T_2$};
      \fill [black] (49.18,-26.82) -- (49.53,-27.7) -- (50.12,-26.89);
      \draw [black] (24.552,-26.406) arc (315.3831:27.3831:2.25);
      \draw (20.03,-24.3) node [left] {$T<T_1$};
      \fill [black] (24.95,-23.79) -- (24.73,-22.87) -- (24.03,-23.58);
      \draw [black] (35.177,-9.52) arc (234:-54:2.25);
      \draw (36.5,-4.95) node [above] {$T<9$};
      \fill [black] (37.82,-9.52) -- (38.7,-9.17) -- (37.89,-8.58);
      \draw [black] (34.6,-14.8) -- (29.07,-23.01);
      \fill [black] (29.07,-23.01) -- (29.94,-22.63) -- (29.11,-22.07);
      \draw [black] (29.09,-23.02) -- (34.81,-14.68);
      \fill [black] (34.81,-14.68) -- (33.94,-15.05) -- (34.77,-15.62);
      \draw [black] (44.7,-23.1) -- (38.3,-14.6);
      \fill [black] (38.3,-14.6) -- (38.38,-15.54) -- (39.18,-14.94);
    \end{tikzpicture}
    \caption{Traffic Light Finite State Machine describes the transition of
      the traffic lights between the three predefined states.}
    \label{fig:TLFSM}
    \vspace{-1.5em}
  \end{center}
\end{figure}

As shown in Algorithm \ref{alg:model_flow}, the final time slot of each cycle is
reserved for the \ac{DM} entity to make a decision. The \ac{DM} analyzes the
information received from the sensors and based on that controls the timing of
the traffic lights. The traffic lights follow a strict set of rules, that
can be summarized with the finite state machine shown in Figure \ref{fig:TLFSM}.
The traffic lights can be configured to be in one of three different states -
State 0, State 1 and State 2. Figure \ref{fig:oneIntersection} shows the system
in State 1, allowing cars traveling eastwards and westwards to pass. In State 0,
all traffic lights are red, and hence no vehicles are allowed to pass through
any traffic lights. State 2 allows only vehicles traveling northbound or
southbound to pass. The \ac{DM} determines how long the traffic lights stay in a
certain state, i.e., the \ac{DM} based on the collected sensor information
calculates the values of $T_1$ and $T_2$ in Figure \ref{fig:TLFSM}. Figure
\ref{fig:TLFSM} also shows that between each transition of State 1 and State 2,
a period of 9 cycles in State 0 takes place. This period allows all traffic that
has recently passed through the traffic lights to safely clear the intersection,
preventing collisions with vehicles coming from other directions.

As previously mentioned, if we consider a more complex scenario (i.e., the
four neighboring intersections scenario), we have to introduce coordination
between the \ac{DM}s. Again, our focus is not the coordination of the decision
making functionalities of the \ac{DM}s in order to optimize the traffic flow.
Therefore, the states of the traffic lights are completely independent from each
other. We focus on the optimization of the communication aspects of this
scenario, meaning that the \ac{DM}s coordinate the transmission time slots for
their sensors in order to minimize the number of collisions. As shown in Figure
\ref{fig:four_intersection_model}, each intersection has 20 sensors, 4 traffic
lights and one \ac{DM}. The vehicles can now be generated in more locations
compared to the basic model (i.e., one intersection model). We also define a
neighbor radius set, that defines the distance between two sensors within which
their transmissions could result in collisions.

In order to coordinate the transmissions for neighboring intersections we
introduce a higher \ac{MAC} layer, that schedules \ac{DM}s in a TDMA like
manner. Each \ac{DM} gets its own dedicated time slot for the communication with
its own sensors. We also introduce a relation between the higher and lower
\ac{MAC} layer time slot lengths. One slot on the higher \ac{MAC} layer is the
equivalent of 20 slots on the lower \ac{MAC} layer. The lower \ac{MAC} layer
protocols are described previously (i.e., slotted Aloha, TDMA and CSMA/CA),
whereas the higher \ac{MAC} layer uses one of the following three approaches to
coordinate the communication amongst multiple \ac{DM}s: (1) Centralized Decision
Maker, (2) Decentralized L-MAC and (3) DESYNC.

\subsection{Centralized Decision Maker}

The approach that involves a \ac{CDM} entity assumes that this centralized node
has all the information needed to control all involved \ac{DM}s. The centralized
node needs to know how many time slots should be assigned to each \ac{DM} and
how to synchronize the activation of all \ac{DM}s. As mentioned previously, our
approach schedules 20 time slots, using a selected protocol - either TDMA,
slotted Aloha or CSMA/CA, per \ac{DM} in a round robin fashion.

\subsection{Decentralized L-MAC}

This approach allows us to coordinate the transmission among multiple \ac{DM}s
by implementing a decentralized TDMA schedule by using the L-MAC protocol
outlined in \cite{fang2013decentralised}. Each \ac{DM} defines a probability
vector of length $C$ where $C$ is the available number of time slots in a round.
Initially, each \ac{DM} chooses a transmission slot with equal probability.
Based on the success and failure rate of transmission for the chosen
transmission slot, the probability vector for each \ac{DM} gets updated to allow
a more intelligent choice of slots in the next round. The result of this is a
collision-free schedule, provided that the number of \ac{DM}s is less than $C$.

\subsection{DESYNC}

The \textsc{Desync} algorithm is described in \cite{degesys2007desync}. Each
\ac{DM} initializes a slot for the communication with its sensor nodes. Each
\ac{DM} also listens for messages that are transmitted by other \ac{DM}s and
stores the timestamps of transmissions that occurred before and after its own
slot. This information is used by the \ac{DM}s to adjust their own slot, by
computing the midpoint between the previous and next slot. The method described
in \cite{degesys2007desync} is concerned with a continuous model. We had to
adapt this in order to fit our discrete model. The midpoint ($T_m$) between the
previous ($T_p$) and next ($T_n$) time slot is computed as,
\begin{equation}
  \label{eq:midpoin}
  T_m = \Bigl\lfloor {\frac{T_p + T_n}{2}} \Bigr\rfloor
\end{equation}
Equation (\ref{eq:midpoin}) was further adapted to deal with the periodic nature
of our timestamps (i.e., time cycles). For example, let us assume that the round
time is $T_r = 10$, $T_p = 8$, and $T_n = 2$. The midpoint slot ($T_m$) calculated for
the next round should be equal to 10. However, using equation (\ref{eq:midpoin})
it results in $T_m = 5$. Therefore, if $T_p$ or $T_n$ is greater than the current
firing slot number, the following equation should be used:

\begin{equation}
  \label{eq:midpoint_modified}
  T_m = \Bigl\lfloor {\frac{T_p + T_n + T_r}{2}} \Bigr\rfloor
\end{equation}
This ensures that each \ac{DM} will position itself in the midpoint slot between
the \ac{DM}s transmitting before and after it, resulting in a collision free
TDMA schedule.


\section{Simulation Study}\label{sec:analysis}
The model was built using the \ac{ABM} \textit{Python} library \textit{Mesa}
\cite{masad2015mesa}. \textit{Mesa} is an open source framework that is built
with the functionality of popular \ac{ABM} simulation software such as NetLogo,
Repast and Mason. \textit{Mesa's} \say{DataCollector} module allows us to
easily collect data from the agents in the model at specified intervals.
\textit{Mesa} enables us to visualize the entire system at each simulation
step, helping with the debugging and verification of the traffic lights
finite state machine.

In this section, we present the simulation results for both scenarios: (1)
uncoordinated and (2) coordinated. The results are generated for a varying range
of input parameters, such as selection of MAC protocols, number of time slots
available for the \ac{DM}s on the higher MAC layer and neighbor radius. The
results are evaluated using the accuracy of information and the spectrum
utilization as criteria.

The accuracy of the information received by the \ac{DM}s is important to the
overall functionality of the system. The actual number of vehicles waiting at a
given moment at the traffic lights is denoted by $N_{W}$. The number of vehicles
that has been registered by the sensor nodes is denoted by $N_{S}$, and the
number of vehicles that has been reported to the \ac{DM} is $N_{DM}$. Since we
focus on the communication aspects of the system, we assume perfect sensing implying $N_S = N_W$. We define accuracy of information as,
\begin{equation}
  \label{eq:aquracy_of_information}
  A = N_{DM} - N_S,
\end{equation}
which implies that the accuracy of information is the difference between the number of
vehicles reported to the \ac{DM} and the actual number of vehicles waiting
at the traffic lights. A positive error ($A > 0$) means that the \ac{DM} believes
that there are more vehicles waiting than the actual figure. A negative error
($A < 0$) means that the \ac{DM} believes that there are less vehicles waiting
than the actual ones. This data is collected once every cycle before the
\ac{DM} action step outlined in Algorithm \ref{alg:model_flow}. Since we assume
perfect sensing ($N_S = N_W$), any discrepancies can be attributed to
interference within the system, i.e., collisions of packets transmitted from
neighboring nodes in the same tick.

The spectrum utilization is a metric that allows us to understand what proportion of
the available information in the system is actually transferred to the \ac{DM}
in order to make a more informed decision about the traffic light states. For
example, if there are 5 vehicles waiting, the total amount of
information/packets that should be available at the \ac{DM} is 5. If 2 sensors
successfully utilize the spectrum, the utilization is 40\%. Therefore,
if the number of successfully transmitted packets in a cycle is denoted with
$N_{succ}$ and the actual number of vehicles waiting at the traffic lights is
$N_{W}$, the spectrum utilization is calculated as:

\begin{equation}
  \label{eq:spectrum_utilization}
  U = \frac{N_{succ}}{N_W}
\end{equation}

\subsection{Uncoordinated Scenario}
\label{sec:uncoordinated_scenario}

The uncoordinated scenario assumes that the \ac{DM}s are not aware of each
other's scheduling decision, meaning that increasing the number of neighboring
intersections will lead to an increase in the number of collisions, due to the
lack of coordination between the neighboring \ac{DM}s.

\begin{figure}[t]
\centering
\includegraphics[scale=0.43]{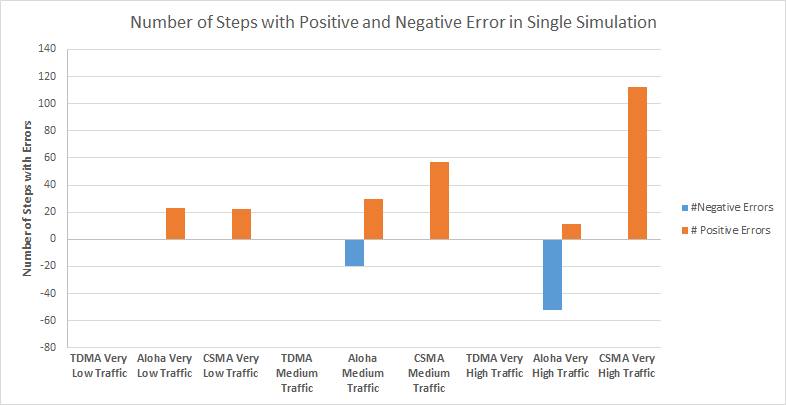}
\caption{Accuracy of Information for the single intersection of roads scenario.
  The neighbor radius is set to 15.}
\label{fig:accuracy}
\vspace{-1.5em}
\end{figure}

Figure \ref{fig:accuracy} shows the average value of the accuracy of information
gathered over $10^4$ simulation steps for the single intersection of roads
scenario. The neighbor radius is set to 15, meaning that sensors that are within
15 hops away can potentially interfere with each other.

Figure \ref{fig:accuracy} highlights that the choice of protocol and level of
traffic affect the accuracy of information received by the \ac{DM}. Regardless
of traffic level, the model using the TDMA protocol make transmissions with zero error,
resulting in highly accurate results. In a single intersection model, each
sensor is allocated its own time slot to send. Therefore, there are no
collisions of packets in a slot, resulting in highly accurate data
transmission to the \ac{DM}. Thus, the \ac{DM} is always aware of exactly how many
vehicles are currently waiting.

When there is low traffic, the model using the slotted Aloha protocol is seen to
have a number of positive errors. The positive errors are due to backed off
sensor packets being transmitted when they no longer reflect the state of the
system (i.e., a collision happens, all involved sensors decide to retransmit the
packets and in the meantime, the traffic lights change state and the number of
vehicles waiting changes). The \ac{DM} is not aware that the received
information is stale and therefore continues to control the timing based on
inaccurate information. The accuracy of information for the slotted Aloha
protocol changes as the number of vehicles waiting grows. This is due to the
fact that with the increasing number of waiting vehicles the number of sensors
trying to transmit their measurements increases. The increased number of
transmissions results in an increasing number of collisions, leading to the case
in which the \ac{DM} does not have the information about a significant number of
vehicles waiting on the lights ($N_{DM} << N_S$).

The model using the CSMA/CA protocol does not suffer from any negative errors.
As expected the number of positive errors increases with the increasing level of
traffic. The reason for this stems from the increased number of sensors
attempting to transmit packets when there is a higher traffic level. When the
traffic lights change state, there is a sudden reduction in the amount of
packets competing for spectrum access as the vehicles begin to move. This leads
to an increased amount of packets reaching the DM with inaccurate information
($N_{DM} >> N_S$).

\begin{figure}[t]
\centering
\includegraphics[scale=0.33]{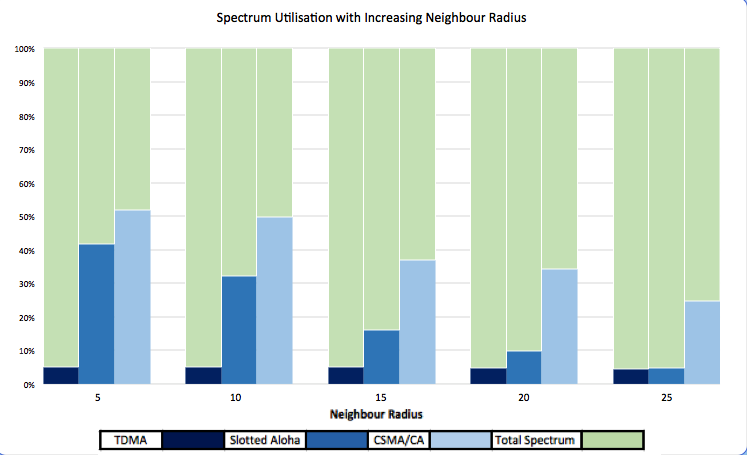}
\caption{Spectrum Utilization - showing how much of the available information in
the system is actually transferred to the DM in order to make a more informed
decision about traffic light system.}
\label{fig:su}
\vspace{-1.5em}
\end{figure}

As previously mentioned, we also calculate the spectrum utilization as shown in
equation (\ref{eq:spectrum_utilization}). The results in Figure \ref{fig:su} are
obtained over $10^4$ simulation steps on a two intersection model. Considering
the uncoordinated nature of the scenario (2 \ac{DM}s that are not aware of each
others scheduling decisions) 40 ticks are assigned for the sensors to transmit
the vehicle detection information per cycle. We up-scaled the model (from one to
two intersections) in order to increase the range of neighboring radii. We vary
the neighbor radius from 5 to 25. All the scenarios assume an intermediate
traffic level, i.e., a 0.5 probability of a car being generated each cycle.

Figure \ref{fig:su}, as expected, shows that TDMA exhibits the lowest spectrum
utilization. However, it is not affected as much by neighbour radius. When
the neighbour radius is large, there is a low probability that two sensors from
different intersections within the same neighbour radius would be scheduled for
the same tick, both having vehicles waiting at them. Hence, the
spectrum utilisation for the TDMA protocol remains fairly constant regardless of
the neighbor radii. CSMA/CA displays the greatest spectrum utilisation in all
variations of neighbor radii. This is due to the \say{sensing first -
  transmitting if available} policy of the CSMA/CA protocol. This allows sensors
to avoid collisions of packets by sensing the collision before it occurs and
backing off for a random period of time. In comparison to this, slotted Aloha
demonstrates a relatively high spectrum utilization when the neighbor radius is
low. The increase of the neighbor radius leads to the increasing number of
collisions (due to lack of coordination and sensing), which results in lower
spectrum utilization.

\subsection{Coordinated Scenario}

The coordinated scenario assumes that the \ac{DM}s are aware of each other's
scheduling. As previously explained, in order to coordinate the scheduling
mechanisms of the \ac{DM}s, we introduce a higher MAC layer. The coordination is
achieved by either a \textit{Centralised Decision Maker},
\textit{\textsc{Desync}} or the \textit{L-MAC} protocol. The operation of the
\textit{Centralised Decision Maker} is obvious - manually configured time slots
for each \ac{DM}. Therefore, we are going to explain in more detail how the
\textsc{Desync} and L-MAC protocols achieve the best time slot assignment on the
higher MAC layer.

\subsubsection{Desync Algorithm}
\begin{figure}
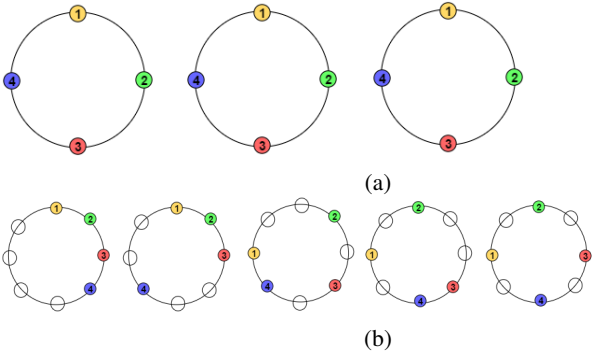

\centering
\begin{subfigure}[b]{0.55\textwidth}
  \includegraphics[scale=0.35]{Figures/DeSyncFour}
   \caption{}
   \label{fig:DeSyncFour}
\end{subfigure}

\begin{subfigure}[b]{0.55\textwidth}
  \includegraphics[scale=0.25]{Figures/DeSync8}
   \caption{}
   \label{fig:DeSync8}
\end{subfigure}

\caption[Two numerical solutions]{\textsc{Desync} (a) T=4  (b) T=8}
\vspace{-1.3em}
\end{figure}

\textsc{Desync} algorithm relies on the fact that each node in the system
performs a task periodically. Depending on the length of the cycle, the
convergence of the system can display different behavior. Figure
\ref{fig:DeSyncFour} depicts a simple scenario in which four \ac{DM}s
coordinate their communication periods. The ring represents the value of $T$,
the total time taken for a full round to be completed. A colored circle with a
number in the middle represents the \ac{DM}, that assigned a time slot. An empty
circle represents a slot where no \ac{DM} is assigned and therefore remains
idle. A round time of four is chosen for this simulation. Each node is given a
unique starting point. The system executes the \textsc{Desync} algorithm.
However, since the nodes are already maximally spaced apart, no further movement
takes place.

A round time of 8 is chosen in the next example. Figure \ref{fig:DeSync8} shows
that all the nodes representing the \ac{DM}s start close to each other in a
round. The \textsc{Desync} algorithm allows the nodes to rearrange themselves
around the ring so that they are maximally spaced apart, and that is when the
system converges (last image in Figure \ref{fig:DeSync8}).

\begin{figure}[t]
\centering
\includegraphics[scale=0.3]{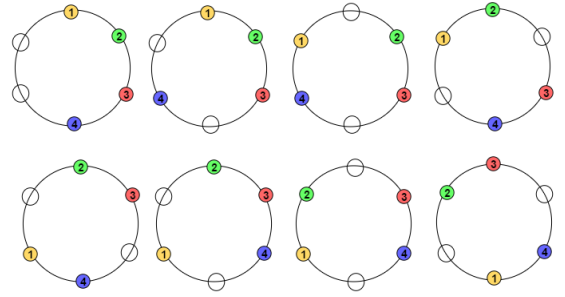}
\caption{\textsc{Desync} T=6}
\label{fig:DeSync6}
\vspace{-1.5em}
\end{figure}

The next example is configured with a round time of 6 (Figure
\ref{fig:DeSync6}). Theoretically, there is no possible configuration to ensure
maximal spacing between the \ac{DM} time slots. Therefore, the system never
converges. Figure \ref{fig:DeSync6} shows that the nodes continue to rearrange
themselves in such a way that a circular pattern emerges. The nodes rotate
counter-clockwise around the ring. This implies that a round time should be chosen such that the number of \ac{DM}s
is a factor of the round time.

\subsubsection{L-MAC Protocol} The method used by the L-MAC protocol to implement a TDMA schedule is quite
different from the \textsc{Desync} protocol. Initially, the \ac{DM}s choose a
random slot in the schedule with equal probability. This is in contrast to the
\textsc{Desync} protocol where nodes are assigned a starting point. If there is
a collision in a slot, the \ac{DM} will choose a slot again in the next round
with updated probabilities. If the \ac{DM} is successful in a slot, it will
choose the same slot again in the next round with a higher probability.

\begin{figure}
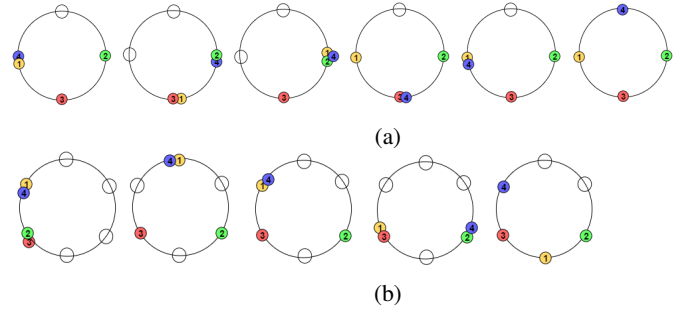

\centering
\begin{subfigure}[b]{0.55\textwidth}
  \includegraphics[scale=0.23]{Figures/LMAC4}
   \caption{}
   \label{fig:LMAC4}
\end{subfigure}

\begin{subfigure}[b]{0.55\textwidth}
  \includegraphics[scale=0.25]{Figures/LMAC6}
   \caption{}
   \label{fig:LMAC6}
\end{subfigure}

\caption[Two numerical solutions]{L-MAC (a) C=4 (b) C=6}
\vspace{-1.5em}
\end{figure}

Figure \ref{fig:LMAC4} shows the convergence of the system to a collision free
configuration. Each node that experiences collisions continues to rearrange
itself, until a collision free schedule is reached. In Figure \ref{fig:LMAC4},
if we take a closer look at \ac{DM}3, we see that due to a collision free
assignment in a previous slot, the node decides to stick with the chosen slot
even after it experiences collisions.



As proven in \cite{fang2013decentralised}, the system can converge with any
round time that is greater than the number of available \ac{DM}s. For the sake
of comparison with the \textsc{Desync} protocol, in Figure \ref{fig:LMAC6}, we
show that the L-MAC protocol can converge with a round time of 6. However, the
L-MAC protocol does not consider the spacing of the nodes around the ring.

Increasing the round time leads to an increasing number of idle slots, which implies that the probability of a node initially choosing slots without collisions
increases as well. That results in a shorter convergence time. Though
very unlikely, it is still possible that the \ac{DM}s randomly choose a collision
free schedule on initial selection with any number of slots in a round greater
or equal to the number of \ac{DM}s.

\begin{figure}[t]
\centering
\includegraphics[scale=0.55]{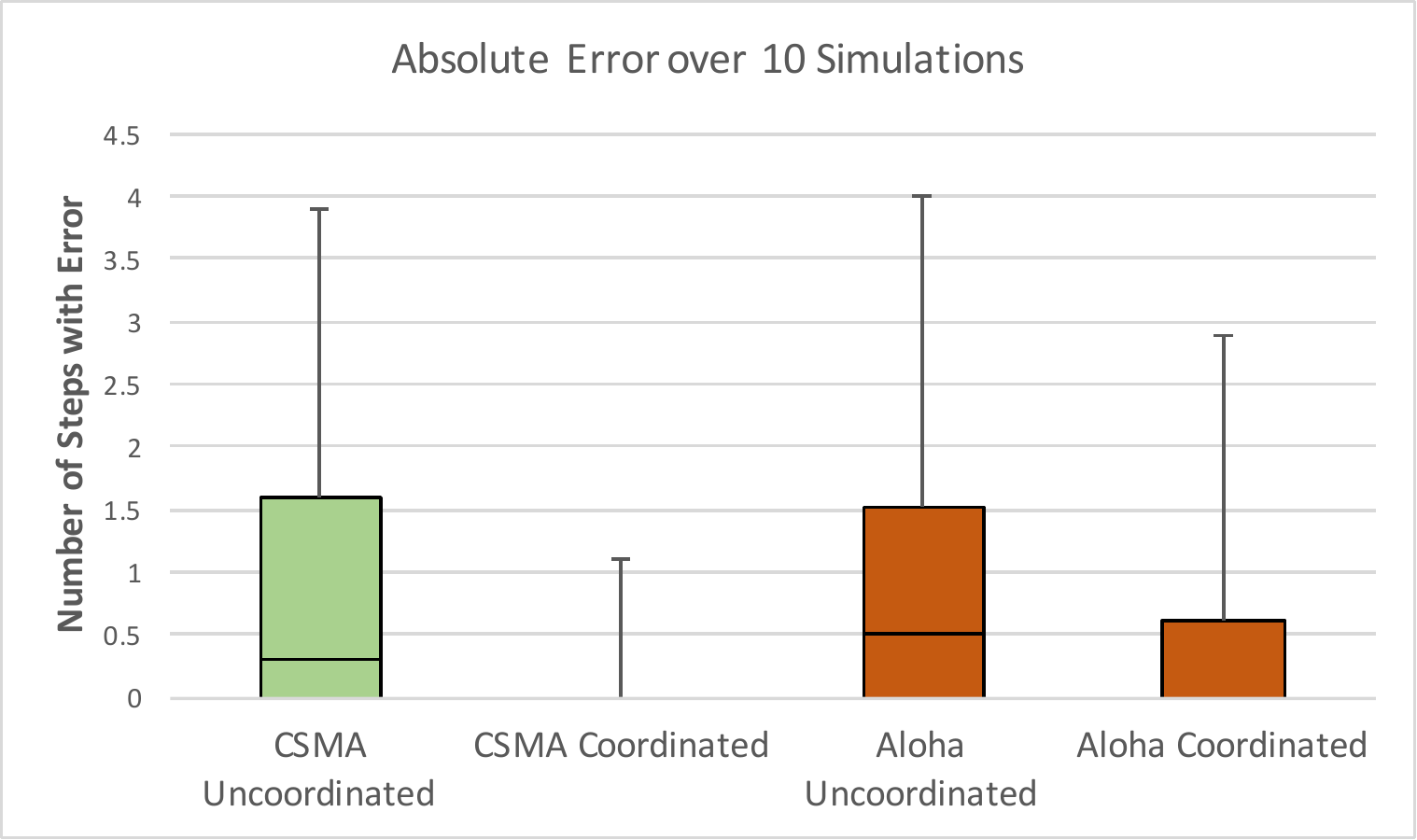}
\caption{Absolute Error - the data is obtained from simulations of the four
  intersection model; traffic level is set to medium; the neighbor radius is set
to 10.}
\label{fig:ae-co}
\vspace{-1.5em}
\end{figure}

Similar to the approach adopted to analyze the uncoordinated scenario, we will
focus on the analysis of the accuracy of information and the spectrum
utilization for the coordinated scenario. The coordinated scenario can be
implemented by using any of the abovementioned high layer MAC protocols. The
number of time slots available on the higher MAC layer is set to four, meaning
that all mentioned higher level MAC protocols would converge to the same
arrangement. We used L-MAC in our simulations. The higher-lower MAC level time
slot length has the ratio of 1:20, meaning that for every higher level MAC time
slot allocated to a \ac{DM}, the equivalent of 20 lower MAC ticks for sensor
transmissions is available. Figure \ref{fig:ae-co} shows the absolute error
(absolute value of the accuracy of information). The data used in Figure
\ref{fig:ae-co} is obtained from simulations of the four intersections model.
The traffic level is set to medium, i.e., a vehicle will be generated with the
probability 0.5 in each cycle. The neighbor radius is set to 10. Figure
\ref{fig:ae-co} depicts the average spread of the absolute error observed over
10 simulations with 5000 simulation steps each. The absolute error is used in
this case as it is not our intention to imply a median error close to zero. The
TDMA results are not shown in Figure \ref{fig:ae-co}, because TDMA results in an
average spread of zero for both uncoordinated and coordinated scenarios.
The data in Figure \ref{fig:ae-co} is shown in the form of a box plot. The upper
extreme of the error bars show the average maximum absolute error, the lower
extreme of the error bar shows the minimum average absolute error. The upper
lines of the boxes in the graph represent the upper quartiles, the lower lines
represents the lower quartiles. The lines in the centers of each box represents
the median values of the absolute error.

Figure \ref{fig:ae-co} shows that the introduction of the higher MAC layer can
greatly reduce the average absolute error for both CSMA/CA and slotted Aloha.
The introduction of coordination reduced the average maximum absolute error from
3.9 to 1.1 for CSMA/CA and from 4 to 2.6 for slotted Aloha. The average median
absolute error was also reduced from 0.3 to 0 for CSMA/CA and from 0.5 to 0 for
slotted Aloha. The reason for the reduction in the absolute error lies in the
influence of the higher MAC layer scheduling procedures. Sensors can only
transmit data to the target \ac{DM} when the target \ac{DM} is selected by the
higher MAC layer. This decreases the problem of stale data. Since the \ac{ABM}
approach allows us to analyze each time step of the discrete simulation, the
analysis of the log files shows that the majority of stale data is received
immediately after a car moving step, mostly affecting the earliest ticks in each
cycle. When a higher MAC layer is introduced, stale data primarily affects the
first \ac{DM} that is selected after a movement step. Previously, all four
\ac{DM}s would be affected by this stale information. The stale information will
indeed only affect the sensors' packets that are being sent to the first \ac{DM}
selected after a movement step, as the selected \ac{DM} will reject all other
sensor packets being transmitted to other \ac{DM}s. Because each \ac{DM} is
allocated its own slot, the interference from sensors transmitting to other
\ac{DM}s is decreased, and thus the accuracy of information received by the
\ac{DM}s is improved.

\begin{figure}[t]
\centering
\includegraphics[scale=0.30]{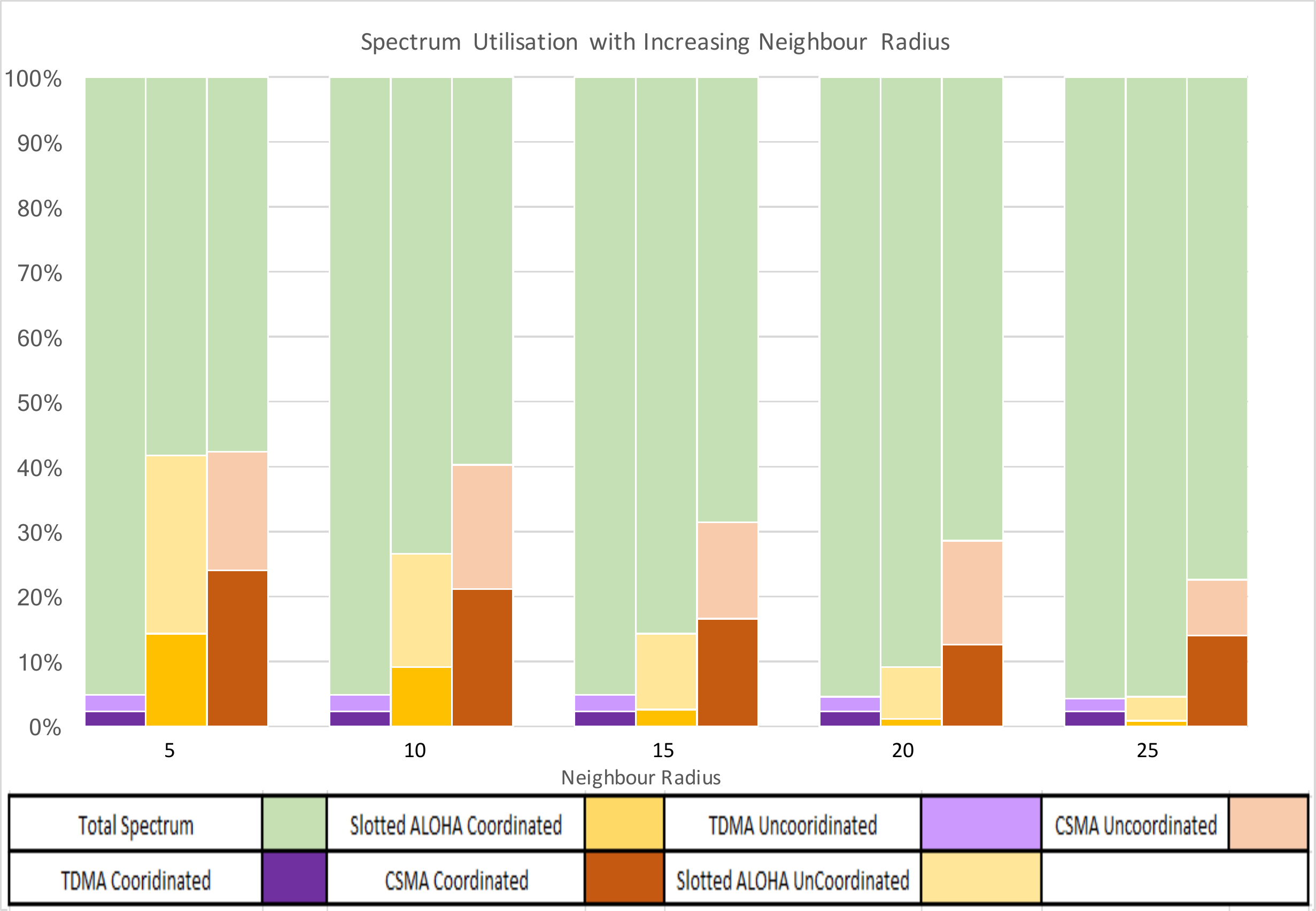}
\caption{Spectrum Utilization - comparison between the coordinated and
  uncoordinated scenario, showing how much of the available information in the
  system is actually transferred to the DM.}
\label{fig:su-co}
\vspace{-1.5em}
\end{figure}

Figure \ref{fig:su-co} shows the comparison between the spectrum utilization for
the coordinated and uncoordinated scenario. The results are gathered from the
simulations of the two intersection model over $10^4$ simulation steps. The
traffic level is again configured to be medium. This is the same configuration
that was used to obtain the spectrum utilization of the uncoordinated model
shown in Figure \ref{fig:su}. The L-MAC protocol is used with a round time of
two, and a higher to lower ratio of 1:20. Therefore, after convergence these
should be no idle slots in the higher MAC layer. This choice of parameters
ensures fairness between the uncoordinated and coordinated scenarios. Two higher
MAC layer ticks in the coordinated scenario with the ratio of 1:20 results in a
total of 40 time slots for the communication between the sensors and the
\ac{DM}s. In the uncoordinated scenario, 40 time slots are assigned for sensor
transmissions in each cycle.

The results obtained from these simulations are overlaid with the results
obtained from the uncoordinated model as shown in Figure \ref{fig:su-co}. It
should be noted that Figure \ref{fig:su-co} is not a stacked bar chart, meaning
that the ratio between each scenario and the total spectrum is being analyzed.
As shown in Figure \ref{fig:su-co}, the spectrum utilisation is reduced when
coordination is introduced. This is due to the limitation that only sensors
transmitting to the selected DM are able to send in each slot. The neighbor
radius has a similar effect on the slotted ALOHA and CSMA/CA protocols in the
uncoordinated and coordinated scenarios (i.e., the spectrum utilization
decreases with increasing neighbor radius). Again TDMA is not affected by the
neighbor radius. As previously explained, in the coordinated TDMA scenario, each
\ac{DM} is assigned its own higher MAC TDMA slot to transmit where each sensor
will then be given its own lower level MAC tick to transmit. The spectrum
utilization of TDMA in the coordinated scenario is approximately half the
spectrum utilization achieved in the uncoordinated scenario. This is due to the
rejection of packets attempting to transmit to \ac{DM}s that are not selected by
the higher MAC layer.


\section{Discussion and Conclusion} \label{sec:conclusion}
The increasing complexity of the next generation of communication networks leads
to a need to change the tools we use to model and analyze them. The primary
purpose of our work is to investigate the possibility of using \ac{ABM} as a
method of modeling an \ac{IoT} network. We show that \ac{ABM} is an effective
way to model the complex behavior of heterogeneous nodes (e.g., simple sensors,
traffic lights, and more powerful decision making nodes). One of the main
advantages of \ac{ABM} is its flexibility in modeling networks that does not
scale up exponentially with the size (e.g., from the simple one intersection
model, to the more complex two and four intersections model). Besides, it
provides opportunity to add new features for decision coordination (e.g., the
higher MAC layer protocol to ensure coordination between the decision makers).
Human interactions (e.g. vehicles) are as easily configurable as the network
agents in the model (e.g. sensors). This feature allowed the level of traffic
and behavior of the vehicles to be modeled as well as the operation of the
network. Another appealing quality of the ABM modeling approach is its ability to model
and collect information on a more granular level (i.e., from all agents within
the system in any time step of the simulation).

In the models where CSMA/CA is used as the selected MAC protocol, we assume
there are no hidden nodes. If hidden nodes were introduced, the behavior and
performance of the models using the CSMA/CA protocol could change. Moreover, we
assumed that the only interference in the model is generated from the agents
within the \ac{IoT} network itself. Although the results in Section
\ref{sec:analysis} suggest TDMA to be an extremely effective MAC protocol, the
limitations of our present model do not highlight the areas where TDMA can fail.
For example, if a sensor fails to transmit successfully due to external source
of interference, it has to wait for its slot in the next cycle to transmit. As
explained in Section \ref{sec:model_description}, the \ac{DM} uses a method of
polling for a certain amount of time before making timing decisions. If this was
more of a continuous decision making process, TDMA could be found to be slower
than the other protocols as each sensor must wait for its time slot and for the
polling phase to be over, before transmitting. Therefore, more investigations
are necessary before concluding in a definite way that TDMA is the best MAC
protocol for the kind of application considered in this paper and a subject of future research.

The motivation for this paper is to investigate \ac{ABM} as a tool for modeling
the complexity of future networks. Building on this work, we envision the future
research to be about modeling of complex networks where ABM can help
to reduce overhead and complexity of distributed decision making.


\begin{acronym}
  \acro{ABM}{Agent-Based Modeling}
  \acro{5G}{Fifth Generation}
  \acro{IoT}{Internet of Things}
  \acro{MAC}{Medium Access Control}
  \acro{CCS}{Complex Communication Science}
  \acro{WSN}{Wireless Sensor Network}
  \acro{DM}{Decision Maker}
  \acro{CDM}{Centralised Decision Maker}
\end{acronym}

%


\ifCLASSOPTIONcaptionsoff
  \newpage
\fi



\bibliographystyle{IEEEtran}
\bibliography{refs}

\end{document}